\documentclass[aps,prl,reprint,superscriptaddress,longbibliography]{revtex4-2}

\usepackage{amsmath,amssymb,bm,physics,mathtools}
\usepackage{graphicx}
\usepackage{xcolor}
\usepackage{hyperref}
\usepackage[T1]{fontenc}
\usepackage[utf8]{inputenc}

\newcommand{\q}{\bm{q}}
\newcommand{\qs}{\q s}
\newcommand{\qsp}{\q s'}

\newcommand{\Hhat}{\hat{H}}

\newcommand{\ii}{\mathrm{i}}

\begin{document}

\title{Lattice thermal transport from phonon spectra beyond perturbation theory}
\author{Zezhu Zeng}
\email{zezhu.zeng@chem.ox.ac.uk}
\affiliation{Physical and Theoretical Chemistry Laboratory, Department of Chemistry, University of Oxford, OX1 3QZ Oxford, UK}

\author{Michele Simoncelli}
\affiliation{Department of Applied Physics and Applied Mathematics, Columbia University, New York 10027, United States}

\author{David E. Manolopoulos}
\affiliation{Physical and Theoretical Chemistry Laboratory, Department of Chemistry, University of Oxford, OX1 3QZ Oxford, UK}

\date{\today}

\begin{abstract}

We develop a molecular dynamics framework to compute the mode-resolved phonon spectral density from classical correlations of an annihilation-like phonon variable. 
For harmonic oscillators, classical molecular dynamics exactly reproduces the corresponding quantum Kubo-transformed correlator, providing the basis for extension to anharmonic systems. 
Using PbTe as a benchmark and Cs$_3$Bi$_2$I$_6$Cl$_3$ as a strongly anharmonic test case, we show that the method captures both quasiparticle and non-Lorentzian spectra beyond perturbative quasiparticle theory, while yielding thermal conductivity in good agreement with experiment. This framework provides a classical molecular dynamics route to mode-resolved phonon spectral densities for spectral Wigner heat transport in strongly anharmonic solids above their Debye temperatures.

\end{abstract}
\maketitle

\section{Introduction}

The theory of lattice thermal transport in crystals originates from the phonon kinetic picture of Peierls \cite{Peierls1929} and the microscopic heat-current formalism of Hardy \cite{Hardy1963}, and is now largely formulated, at the first-principles level, in a harmonic phonon basis dressed by anharmonic interactions within perturbation theory (PT) \cite{Broido2007,Fugallo2013,Barbalinardo2020kALDo,Cahill2014NanoscaleII,lindsay2018survey,Togo2023JPSJ,Togo2023JPCM}. 
In weakly anharmonic crystals, this framework is successful: harmonic interatomic force constants (IFCs) define the phonon eigenstates, anharmonic IFCs determine phonon linewidths, and the lattice thermal conductivity ($\kappa$) follows from the phonon Boltzmann equation.
Recently, this framework has been generalized, relying on the Wigner \cite{Simoncelli2019,Simoncelli2022} or quasi-harmonic Green-Kubo formalisms \cite{Isaeva2019,Fiorentino2023}, into a unified formulation that encompasses both the Peierls limit for crystals and the Allen-Feldman \cite{AllenFeldman1989} limit for glasses, covering also the intermediate cases.

This description becomes increasingly fragile, however, in strongly anharmonic materials, where large frequency renormalization, severe phonon scattering, and eventually overdamped vibrational dynamics render the very notions of a single phonon frequency and lifetime progressively ill defined \cite{dangicOriginLatticeThermal2021a,Wang2025Overdamped,Dangic2025Overdamped}.  In this regime, the relevant object is no longer a scalar lifetime $\tau_{\mathbf q s}$, but rather the full mode-resolved spectral density $b_{\mathbf q s}(\omega)$ associated with the retarded phonon self-energy $\Sigma_{\mathbf q s}^{R}(\omega)=\Delta_{\mathbf q s}(\omega)-i\Gamma_{\mathbf q s}(\omega)$.
Crucially, Caldarelli \emph{et al.} \cite{Caldarelli2022} have shown that this spectral density can be inserted directly into an integrated Wigner transport framework, so that $\kappa$ remains well defined even beyond the Lorentzian quasiparticle limit.

Existing spectral constructions, however, remain largely tied to perturbative self-energies, whether at the bubble level \cite{Togo2023JPSJ,Togo2023JPCM} or in more elaborate self-consistent extensions \cite{Paulatto2015PdH,Tadano2022HalidePerovskite,Xiao2023,Monacelli2025,Xia2025}. The central challenge is therefore how to obtain $b_{\mathbf q s}(\omega)$ reliably once finite-order perturbation theory becomes insufficient. In this work we formulate a direct route to $b_{\mathbf q s}(\omega)$ from molecular dynamics (MD). Rather than reconstructing the spectrum from a chosen self-energy approximation, we project finite-temperature trajectories onto normal-mode coordinates \cite{Ladd1986PRB,McGaughey2014PredictingPhonons}, compute the time correlation of the complex phonon amplitude, and Fourier transform it to obtain the one-phonon $b_{\qs}(\omega)$. This can then be inserted directly into the spectral Wigner conductivity expression\cite{Caldarelli2022} without assuming either a Lorentzian line shape or an explicit truncation of the self-energy,  enabling a direct MD-based spectral interpretation of heat transport in terms of both  particle-like intraband propagation and wave-like interband tunneling of phonons \cite{Simoncelli2022}.

\section{Formalism}

We begin with the spectral Wigner expression for the thermal conductivity tensor \cite{Caldarelli2022},
\begin{align}
\kappa^{\alpha\beta}
&=
\frac{\pi\hbar^2}{N_{\q}Vk_BT^2}
\sum_{\q,ss'}
\frac{\left(\omega_{\qs}+\omega_{\qsp}\right)^2}{4}\,
{\sf v}^{\alpha}_{ss'}(\q)\,
{\sf v}^{\beta}_{s's}(\q)
\nonumber \\
&\quad\times
\int_0^{\infty} {\rm d}\omega\,
b_{\qs}(\omega)\,
b_{\qsp}(\omega)\,n(\omega,T)[n(\omega,T)+1],
\label{eq:kappa_wigner_prb2022}
\end{align}
where \(N_{\q}\) is the number of sampled wave vectors, $V$ is the unit-cell volume, ${\sf v}^\alpha_{ss'}(\q)$ is a Wigner velocity matrix element, \(b_{\qs}(\omega)\) is the spectral density of a phonon mode with wave vector $\q$ and branch index $s$, and $n(\omega,T)=(e^{\hbar\omega/k_BT}-1)^{-1}$ is a Bose-Einstein occupation number. In the weak-damping limit, Eq.~\eqref{eq:kappa_wigner_prb2022} reduces to the standard Wigner transport equation \cite{Simoncelli2019} for phonons with Lorentzian lineshapes, but it remains valid for more general spectral densities.

The central quantity is therefore \(b_{\qs}(\omega)\). 
This can be related to the Fourier transform of the (greater) real-time quantum correlator \(c^{\rm Q}_{\qs}(t)=\langle \hat a_{\qs}(t)\hat a_{\qs}^{\dagger}(0)\rangle\)~\cite{Caldarelli2022} using a detailed balance relation (see Appendix A for the detailed derivation):
\begin{equation}
b_{\qs}(\omega) = {c^{\rm Q}_{\qs}(\omega)\over 2\pi[n(\omega,T)+1]}
\label{eq:CQ_b},
\end{equation}
where
\begin{equation}
c^{\rm Q}_{\qs}(\omega) =
\int_{-\infty}^{\infty}{\rm d}t\,e^{+\ii\omega t}c^{\rm Q}_{\qs}(t).
\label{eq:CQ_def}
\end{equation}
The operator \(\hat a_{\qs}(t)\) in the correlator $c^{\rm Q}_{\qs}(t)$ is the time-evolved annihilation operator of phonon mode $\qs$. In a harmonic phonon basis, this can be constructed as $\hat a_{\qs}(t) = \sqrt{\omega_{\qs}/2\hbar}\,\hat u_{\qs}(t)+\ii/\sqrt{2\hbar\omega_{\qs}}\,\hat p_{\qs}(t)$, where $\hat{u}_{\qs}(t)$ and $\hat{p}_{\qs}(t)$ are the mode-projected atomic displacements and momenta 
\begin{align}
\hat{u}_{\qs}(t)
&=
\frac{1}{\sqrt{N_{\q}}}
\sum_{lb\alpha}
\sqrt{M_b}\,\hat{u}_{lb\alpha}(t)\,
e^{-\ii\q\cdot(\bm R_l+\bm\tau_b)}
{\cal E}^*(\q)_{s,b\alpha}
\nonumber\\
\hat{p}_{\qs}(t)
&=
\frac{1}{\sqrt{N_{\q}}}
\sum_{lb\alpha}
\frac{\hat{p}_{lb\alpha}(t)}{\sqrt{M_b}}\,
e^{-\ii\q\cdot(\bm R_l+\bm\tau_b)}
{\cal E}^*(\q)_{s,b\alpha}.
\label{eq:up}
\end{align}
Here ${\bm R}_l$ is a lattice vector, $M_b$ and $\bm{\tau}_b$ are the mass and  equilibrium position of atom $b$ within the unit-cell, and ${\cal E}(\q)$ is the eigenvector matrix of the dynamical matrix at wave vector $\q$ \cite{Caldarelli2022}.

To make a connection with molecular dynamics, we introduce the Kubo correlator
\begin{equation}
c^{\rm K}_{\qs}(t) =
k_BT\int_{0}^{1/k_BT}d\lambda\,
\left\langle
\hat a_{\qs}(t-\ii\lambda\hbar)\hat a_{\qs}^{\dagger}(0)
\right\rangle,
\label{eq:cK_def}
\end{equation}
and use the identity \cite{Kubo1957,Craig2004}
\begin{equation}
c^{\rm K}_{\qs}(\omega) =
\int_{-\infty}^{\infty}{\rm d}t\,e^{+\ii\omega t} c^{\rm K}_{\qs}(t)
= {k_BT\over\hbar\omega}{c^{\rm Q}_{\qs}(\omega)\over [n(\omega,T)+1]}
\label{eq:detailed-balance}
\end{equation}
to re-write Eq.~\eqref{eq:CQ_b} as
\begin{equation}
b_{\qs}(\omega) =
\frac{\hbar\omega}{2\pi k_BT}\,
c^{\rm K}_{\qs}(\omega).
\label{eq:b_from_K}
\end{equation}
This can be obtained directly from molecular dynamics trajectories by making the approximation~\cite{Sun2010Allen} 
\begin{equation}
b_{\qs}(\omega) \approx
\frac{\hbar\omega}{2\pi k_BT}\,
c_{\qs}^{\rm CL}(\omega),
\label{eq:b_from_MD}
\end{equation}
where $c_{\qs}^{\rm CL}(\omega)$ is the Fourier transform of the classical correlator $c_{\qs}^{\rm CL}(t)=\langle a_{\qs}(t)a^{*}_{\qs}(0)\rangle$ of the annihilation-like variable $a_{\qs}(t) = \sqrt{\omega_{\qs}/2\hbar}\,u_{\qs}(t)+\ii/\sqrt{2\hbar\omega_{\qs}}\,p_{\qs}(t)$. 

The replacement of a Kubo correlator with a classical correlation function will be familiar to the phonon community~\cite{Ladd1986PRB,Sun2010Allen,McGaughey2014PredictingPhonons,Lv2016GKMA}. It should be understood as a classical-limit approximation to quantum real-time dynamics: it is exact in the harmonic limit because the annihilation operator is linear in displacements and momenta (see the SI), and it becomes exact in the high-temperature limit where $k_{B}T\gg\hbar\omega_{\qs}$  for all $\qs$ regardless of the strength of the anharmonicity. Unlike the (greater) quantum correlator $c_{\qs}^{\rm Q}(t)$, the Kubo correlator $c_{\qs}^{\rm K}(t)$ has the same time-reversal and detailed balance symmetries as the classical correlation function $c_{\qs}^{\rm CL}(t)$ \cite{Craig2004}, and expressions for linear response observables have the same form when written quantum mechanically in terms of $c^{\rm K}_{\qs}(t)$ as they do when written classically in terms of $c_{\qs}^{\rm CL}(t)$ \cite{Kubo1957}.

The approximation in Eq.~(8) will of course break down for the phonons with $\hbar\omega_{\qs}\gg k_{\rm B}T$ in light-atom materials at low temperatures, and there is little that can be done about this because the non-perturbative evaluation of real-time quantum correlation functions in this regime is still an open and highly challenging problem. Here we shall therefore focus on heavy-atom materials above their Debye temperatures, for which Eq.~(8) is best justified, and show that it provides a practical route from classical MD to Wigner transport in situations where the phonon spectra are non-Lorentzian and the phonon lifetimes are ill-defined.

\begin{figure*}[t]
\centering
\includegraphics[width=1\textwidth]{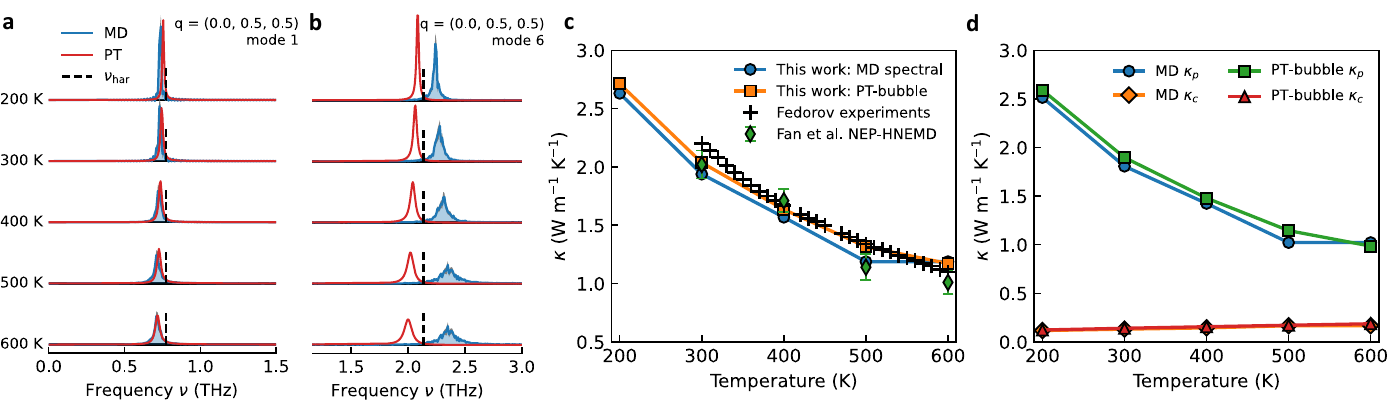}
\caption{Mode-resolved phonon spectral densities $b_{\mathbf q s}(\omega)$ of PbTe at $\mathbf q=(0,1/2,1/2)$ for (a) the lowest-frequency mode and (b) the highest-frequency mode, from 200 to 600~K.  
Blue lines show the averaged MD $b_{\mathbf q s}(\omega)$ over 20 independent trajectories, with gray shading denoting the standard error. Red lines show the perturbative bubble results, and black dashed lines mark the harmonic reference frequencies.
(c) Total lattice thermal conductivity $\kappa$ as a function of temperature, comparing the present MD-spectral and perturbative-bubble results with experiment \cite{Fedorov1969PbTe} and previous homogeneous non-equilibrium MD calculations \cite{Fan2021NEP}. 
(d) Decomposition of the MD and perturbative-bubble conductivities into intraband propagation ($\kappa_{\mathrm P}$, $s=s'$) and interband tunneling ($\kappa_{\mathrm C}$, $s\not=s'$) contributions.}
\label{fig:pbte_overview}
\end{figure*}

\section{P\MakeLowercase{b}T\MakeLowercase{e}: Weak-to-strong anharmonic crossover}

PbTe serves as a benchmark system that spans weak-to-strong anharmonicity from 200 to 600~K \cite{Delaire2011PbTe,Ribeiro2018PbTeSnTe,Shiga2012PbTe}. 
It is therefore an ideal first test of the present framework: at room temperature, perturbative phonon self-energy theory is expected to remain broadly reliable, whereas at higher temperature higher-order anharmonic renormalization is expected to become increasingly important. The computational details are provided in SI.

We first examine the mode-resolved spectral densities for the lowest-frequency mode at $\mathbf q=(0,1/2,1/2)$.  Fig.~\ref{fig:pbte_overview}(a) shows that the MD spectral density of this mode remains in very good agreement with the bubble result over the full temperature range. This provides an important validation of the present method, showing that the projection onto normal modes, the extraction of the time correlator, and the reconstruction of $b_{\mathbf q s}(\omega)$ recover the expected perturbative limit in a weakly anharmonic regime. 
By contrast, for the highest-frequency mode shown in Fig.~\ref{fig:pbte_overview}(b), the MD spectrum develops a clear split structure around the harmonic frequency that is absent in the bubble result, although the overall lineshape remains broadly quasiparticle-like.
For the high-frequency optical mode, the MD spectrum is broader and blue-shifted relative to the lowest-order bubble result, with a peak position consistent with the measurement of Cochran \textit{et al.} ($2.35~\mathrm{THz}$)~\cite{Cochran1966PbTe}. 
This trend is also consistent with the first-principles studies of Xia~\cite{Xia2018PbTeQuartic,Xia2025}, which have shown that fourth-order anharmonicity plays an important role in hardening the optical phonons of PbTe.
In this sense, PbTe illustrates the advantage of the MD route: it recovers the perturbative limit where expected, while remaining sensitive to spectral features in $b_{\mathbf q s}(\omega)$ that are beyond the lowest-order of perturbation theory.

The same conclusion is reflected in $\kappa$. 
Fig.~\ref{fig:pbte_overview}(c) shows that the conductivities obtained from the MD and perturbative spectral densities are in close agreement. Both are consistent with experiment \cite{Fedorov1969PbTe}, and with previous homogeneous non-equilibrium MD simulations \cite{Fan2021NEP}.
The MD conductivity is, however, slightly lower than that of perturbation theory. As shown in Fig.~\ref{fig:pbte_overview}(d), the reduction is in the intraband propagation contribution $\kappa_{\mathrm P}$, while the interband tunneling contribution $\kappa_{\mathrm C}$ is very similar in the two cases. 
The reduced $\kappa_{\rm P}$ is consistent with the stronger broadening of optical modes in the MD spectral density, reflecting full-order anharmonic renormalization beyond the lowest-order bubble approximation.

\begin{figure*}[t]
\centering
\includegraphics[width=1\textwidth]{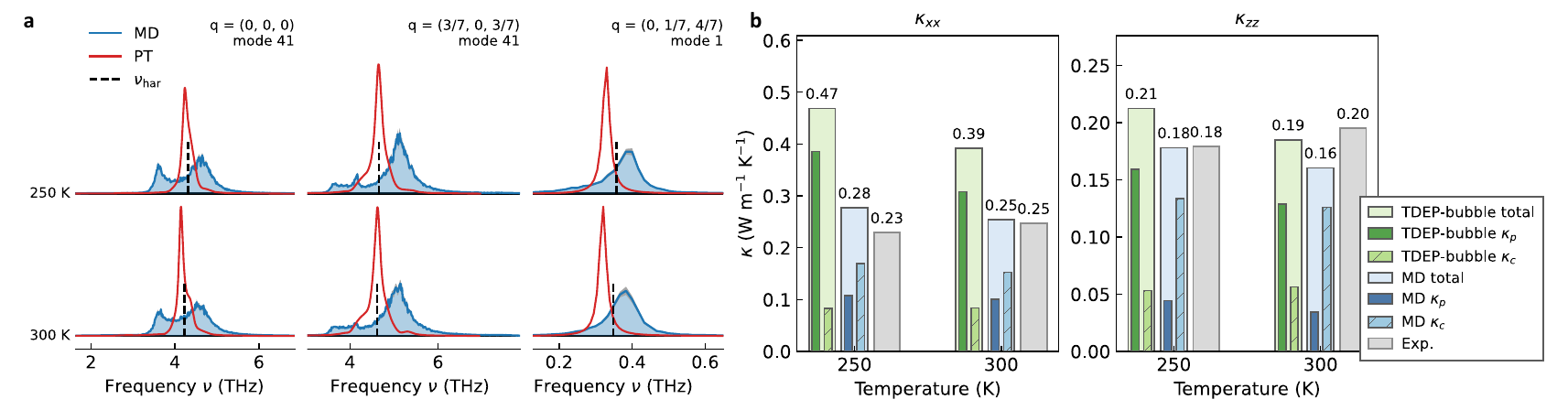}
\caption{(a) Representative mode-resolved phonon spectral densities $b_{\mathbf q s}(\omega)$ of Cs$_3$Bi$_2$I$_6$Cl$_3$ with various wave vectors and branches, at 250 and 300~K.
Blue lines and corresponding shaded gray regions show the averaged MD $b_{\mathbf q s}(\omega)$ and the standard error over 48 independent trajectories, respectively; blue lines denote the TDEP-bubble $b_{\mathbf q s}(\omega)$, while black dashed lines mark the harmonic reference frequencies.
(b) Thermal conductivities along the $x$ and $z$ directions, comparing the total conductivity and its intraband propagation ($\kappa_{\mathrm P}$) and interband tunneling ($\kappa_{\mathrm C}$) contributions obtained from the MD and TDEP-bubble spectral densities, together with experiment \cite{Zeng2025PNAS}. 
}
\label{fig:csbi_spectral}
\end{figure*}

\section{C\MakeLowercase{s}$_3$B\MakeLowercase{i}$_2$I$_6$C\MakeLowercase{l}$_3$: Phonons beyond quasiparticles}

While PbTe provides a  benchmark in a regime where PT remains meaningful, Cs$_3$Bi$_2$I$_6$Cl$_3$ offers a more stringent test in which the quasiparticle picture itself becomes fragile. This material is especially pertinent because our recent work \cite{Zeng2025PNAS} has shown that its $\kappa$ is exceptionally low, while perturbative calculations, even after including four-phonon scattering \cite{Feng2016FourPhonon,Feng2017FourPhonon}, still tend to overestimate the experimental conductivity. 
At the same time, Cs$_3$Bi$_2$I$_6$Cl$_3$ is not well described by a simple crystalline reference at $T=0$: it exhibits lattice distortion at low temperature, whereas thermal fluctuations restore the average $P6_3/mmc$ phase above roughly 200~K \cite{Zeng2025PNAS}. 
The harmonic phonons of the average crystal structure therefore contain imaginary modes and are not an appropriate basis for the present analysis. 
For this reason, the mode projection and transport calculation reported below are formulated in a temperature-renormalized phonon basis obtained from a temperature-dependent effective potential (TDEP) calculation \cite{Hellman2011TDEP,Hellman2013TDEP}.
Details are provided in the SI.

After establishing the finite-temperature $P6_3/mmc$ phonon basis  at 250 and 300~K, we examine the mode-resolved spectral densities in Fig.~\ref{fig:csbi_spectral}(a). 
In contrast to PbTe, Cs$_3$Bi$_2$I$_6$Cl$_3$ already shows pronounced non-Lorentzian behavior at the spectral level. 
The TDEP-bubble results feature only modest deviations from a simple Lorentzian profile, whereas the MD spectra display much richer lineshapes. 
The three representative examples in Fig.~\ref{fig:csbi_spectral}(a) highlight distinct manifestations of this effect: a double-peak structure for a high-frequency optical mode at the $\Gamma$ point, a strongly lopsided and broadly distributed optical spectrum at $\mathbf q=(3/7,0,3/7)$, and an asymmetric low-frequency acoustic peak at $\mathbf q=(0,1/7,4/7)$. These features are far more pronounced than in the perturbative spectra and indicate that the vibrational dynamics can no longer be reduced to a single renormalized frequency and linewidth. Instead, the MD spectral density captures higher-order anharmonic self-energy effects directly, including substantial spectral redistribution, broadening, and mode-shape distortion beyond the reach of the TDEP-bubble description.

These spectral differences translate directly into the conductivities shown in Fig.~\ref{fig:csbi_spectral}(b). 
Along the $x$ direction, the TDEP-bubble approach  overestimates $\kappa^{xx}$ at both 250 and 300~K; in contrast, the MD-based result is in much better agreement with experiment \cite{Zeng2025PNAS}. 
The origin of this improvement is apparent from the decomposition into intraband propagation and interband tunneling contributions: the stronger spectral broadening in MD leads to a substantial suppression of $\kappa_{\mathrm P}$. 
Along the $z$ direction, both TDEP and MD give total conductivities close to experiment, especially at 300~K, but the underlying transport mechanism changes qualitatively. In the TDEP-bubble picture, $\kappa^{zz}$ remains propagation-dominated, while in the MD-based description the tunneling contribution is larger than the propagation one. This crossover is a direct consequence of the broadened and non-Lorentzian spectral densities obtained from MD, which enhance spectral overlap between modes and therefore strengthen interband tunneling transport mechanisms.
Cs$_3$Bi$_2$I$_6$Cl$_3$ thus provides the paradigmatic example in which the full spectral shape controls the resulting heat transport.

\section{Discussion}

Our MD framework is designed to preserve the spectral and mode-resolved language of phonon transport while replacing a finite-order self-energy construction. 
This can be distinguished from conventional normal-mode analysis (NMA) approaches~\cite{Ladd1986PRB,McGaughey2014PredictingPhonons} and Green-Kubo modal analysis (GKMA)~\cite{Lv2016GKMA}. 
All these methods analyze classical MD trajectories, but for constructing different transport quantities. 
NMA extracts renormalized phonon frequencies and lifetimes by fitting Lorentzian-like mode spectra, whereas GKMA decomposes the real-space MD heat current into modal contributions of phonons. 
In contrast, we focus on the full phonon spectral density and insert it into the Wigner spectral transport~\cite{Caldarelli2022}, allowing non-Lorentzian spectra and their overlaps to enter $\kappa$ without reducing each mode to a single lifetime.

Our method becomes particularly valuable in strongly anharmonic systems, where perturbative treatments face both physical and practical limitations. 
Physically, high-order phonon scattering processes can make significant contributions to the spectral shape. 
Practically, the extraction of high-order IFCs is costly and uncertain, cutoff choices for IFCs can considerably affect the results, and the numerical treatment of energy conservation in phonon-phonon processes, for example through Gaussian smearing, introduces additional ambiguity \cite{Castellano2025FDT,dilucentePhononCollisionalBroadening2026}. 
In contrast, the MD route absorbs these anharmonic  effects completely into the spectral density, without explicit enumeration of bubble \cite{Tadano2022HalidePerovskite}, tadpole \cite{Monacelli2025}, loop \cite{Tadano2015SCPH}, sunset \cite{Xiao2023}, even fifth- or sixth-order phonon processes \cite{Xia2025FiveSix}.

A further advantage is that, in addition to the MD trajectory, the present method requires only an effective second-order phonon basis for mode projection and transport. 
This makes it especially attractive for structurally complex crystals with many atoms per unit cell, such as the metal-organic framework MOF-5 \cite{Huang2007MOF5MD} and the Zintl thermoelectric Yb$_{14}$MnSb$_{11}$ \cite{Wang2018Yb14MnSb11}. 
In these cases, high-order IFC approaches become very expensive, whereas the present approach makes it practical to compute and microscopically interpret the thermal conductivity.
More generally, this approach enables one to resolve the microscopic heat-conduction mechanisms in any situation where a MD simulation is feasible and an average crystal structure (and hence the second-order phonon basis) can be defined \cite{simoncelliTemperatureinvariantCrystalGlass2025,iwanowskiBondNetworkEntropyGoverns2025,thebaudBreakingRayleighsLaw2023,fiorentinoEffectsColoredDisorder2025,pazhedathFirstprinciplesCharacterizationThermal2024,kotiugaMicroscopicPictureParaelectric2022a}.
We also stress that the present spectral Wigner implementation uses the harmonic heat-flux~\cite{Hardy1963}. 
Anharmonicity enters through the MD time evolution and the resulting interacting spectral density, while explicit higher-order anharmonic heat-flux is not included.

\section{Conclusions}

We have developed a MD framework to compute mode-resolved phonon spectral densities from classical correlations of a quantum annihilation-like phonon variable and to insert them directly into the spectral Wigner formulation of lattice thermal transport. 
We have shown that the method recovers the perturbative limit at low temperature while capturing higher-order anharmonic corrections at elevated temperature. 
Moreover, recent machine-learning potentials such as NEP~\cite{Fan2021NEP}, together with GPU-accelerated MD engines such as GPUMD~\cite{Fan2022GPUMD}, make statistically converged anharmonic trajectories increasingly feasible for complex materials.
The present framework thus provides a practical methodology that connects classical MD to spectral Wigner transport, enabling us to resolve the microscopic heat transport mechanisms underlying the macroscopic thermal conductivity.

\appendix
\setcounter{equation}{0}
\renewcommand{\theequation}{A\arabic{equation}}

\section*{Appendix A: Derivation of Eq.~(2)}
\addcontentsline{toc}{section}{Appendix A: Derivation of Eq.~(2)}

In this paper, $b_{\qs}(\omega)$ is obtained from the real-time correlator $c^{\rm Q}_{\qs}(t)$ rather than from a time-ordered Green's function. This is not a different definition of the spectral density: it is the same $b_{\qs}(\omega)$ as in the Matsubara/retarded formulation of Caldarelli {\em et al.}~\cite{Caldarelli2022}, re-expressed using a detailed balance relation.

Caldarelli {\em et al}.~\cite{Caldarelli2022} used a standard procedure to define their phonon spectral density $b_{\qs}(\omega)$. They started from the time-ordered imaginary-time correlator
\begin{equation}
\begin{aligned}
g_{\qs}(\tau) &= -\left\langle{\cal T}_{\tau}\hat a_{\qs}(\tau)\hat a^{\dagger}_{\qs}(0) \right\rangle \\
\hat a_{\qs}(\tau) &=e^{+\Hhat\tau}\hat a_{\qs}e^{-\Hhat\tau}
\qquad 0\le\tau\le\beta,
\end{aligned}
\end{equation}
which they expanded in the Fourier series 
\begin{equation}
\begin{aligned}
g_{\qs}(\tau) &= \frac{1}{\beta}\sum_{l=-\infty}^{\infty} e^{-\ii\Omega_l\tau}g_{\qs}(\ii\Omega_l)\\
g_{\qs}(\ii\Omega_l) &= \int_0^{\beta} {\rm d}\tau\, e^{+\ii\Omega_l\tau}g_{\qs}(\tau),
\end{aligned}
\end{equation}
where $\Omega_l=2\pi l/\beta$ with $\beta=1/k_{\rm B}T$ are bosonic Matsubara energies (energies rather than frequencies because we are following their notation \cite{Caldarelli2022}). The Fourier coefficients $g_{\qs}(\ii\Omega_l)$ in this expansion have the Stieltjes spectral representation
\begin{equation}
g_{\qs}(\ii \Omega_l) = \int_{-\infty}^{+\infty}\dd\omega\,\frac{b_{\qs}(\omega)}{\ii\Omega_l-\hbar\omega},
\end{equation}
which can be analytically continued to give the Green's function
\begin{equation}
g_{\qs}(z) = \int_{-\infty}^{+\infty}\dd\omega\,\frac{b_{\qs}(\omega)}{z-\hbar\omega}
\end{equation}
for $z\in \mathbb{C} \setminus \mathbb{R}$. The boundary value of $g_{\qs}(z)$ just above the real axis can be evaluated using the Sokhotski-Plemelj identity to give
\begin{equation}
g_{\qs}(\hbar\omega+\ii 0^+) = {\cal P}\int_{-\infty}^{\infty} {\rm d}\omega' \frac{b_{\qs}(\omega')}{\hbar(\omega-\omega')}-\ii\frac{\pi}{\hbar}b_{\qs}(\omega),
\end{equation}
and therefore
\begin{equation}
    b_{\qs}(\omega) = -\frac{\hbar}{\pi} {\rm Im}\,g_{\qs}(\hbar\omega+\ii 0^+),
\end{equation}
which is Eq.~(38) in Caldarelli {\em et al.}'s paper \cite{Caldarelli2022}.

To connect this Matsubara formulation to our real-time correlator $c^{\rm Q}_{\qs}(t)$, we will begin by evaluating $g_{\qs}(\tau)$ in the eigenbasis of the Hamiltonian. (Note that the harmonic phonon basis enters only through the projected operator $\hat a_{\qs}$ -- the dynamics carried by the eigenstates of $\hat{H}$ is fully anharmonic -- and that since $\tau$ is in the range $0\le\tau\le\beta$ the imaginary time-ordering operator ${\cal T}_{\tau}$ has no role to play.) Re-writing Eq.~(A1) as
\begin{equation}
    \begin{aligned}
        g_{\qs}(\tau) &= -\frac{1}{Z}{\rm tr}\left[e^{-\beta\hat{H}}\hat{a}_{\qs}(\tau)\hat{a}_{\qs}^{\dagger}(0)\right]\\
        Z &= {\rm tr}\left[e^{-\beta\hat{H}}\right], 
    \end{aligned}
\end{equation}
and inserting two complete sets of energy eigenstates 
$1 = \sum_{m} |m\rangle\langle m|$ gives 
\begin{equation}
g_{\qs}(\tau) = -\frac{1}{Z}\sum_{mn} e^{-\beta E_m}e^{\tau(E_m-E_n)}|\langle m|\hat{a}_{\qs}|n\rangle|^2. 
\end{equation}
Substituting this into Eq.~(A2) and evaluating the integral over $\tau$ gives
\begin{equation}
g_{\qs}(\ii \Omega_l) = \frac{1}{Z}\sum_{mn} \frac{e^{-\beta E_m}-e^{-\beta E_n}}{\ii\Omega_l-(E_n-E_m)}|\langle m|\hat{a}_{\qs}|n\rangle|^2,
\end{equation}
and therefore, by comparison with Eq.~(A3),
\begin{equation}
\begin{aligned}
b_{\qs}(\omega) = \frac{1}{Z}\sum_{mn} &\left(e^{-\beta E_m}-e^{-\beta E_n}\right)|\langle m|\hat{a}_{\qs}|n\rangle|^2\\
&\times \delta(\omega-[E_n-E_m]/\hbar),
\end{aligned}
\end{equation}
which is the Lehmann representation of Caldarelli {\em et al.}'s spectral density.

The real-time correlator $c^{\rm Q}_{\qs}(t)=\langle \hat{a}_{\qs}(t)\hat{a}^{\dagger}(0)\rangle$, by contrast, involves just a single Boltzmann weight, because its operators are in a fixed order (without any time ordering, step function, or commutator):
\begin{equation}
c^{\rm Q}_{\qs}(t) = \frac{1}{Z}\sum_{mn}e^{-\beta E_m}e^{-\ii(E_n-E_m)t/\hbar}\left|\langle m|\hat a_{\qs}|n\rangle\right|^2.
\end{equation}
For a transition with $\hbar\omega=E_n-E_m$, detailed balance gives
$e^{-\beta E_n}=e^{-\beta E_m}e^{-\beta\hbar\omega}$, and therefore
\begin{equation}
e^{-\beta E_m}
=
\left[n(\omega,T)+1\right]
\left(e^{-\beta E_m}-e^{-\beta E_n}\right),
\label{eq:app_detailed_balance}
\end{equation}
where $n(\omega,T)=(e^{\beta\hbar\omega}-1)^{-1}$. Inserting this relation into Eq.~(A11) and using Eq.~(A10) gives
\begin{equation}
c^{\rm Q}_{\qs}(t)
=
\int_{-\infty}^{+\infty}\dd\omega\,
e^{-\ii\omega t}
\left[n(\omega,T)+1\right]b_{\qs}(\omega).
\label{eq:app_fdt_time}
\end{equation}
With the Fourier convention
$c^{\rm Q}_{\qs}(\omega)=\int_{-\infty}^{\infty}\dd t\,e^{+\ii\omega t}c^{\rm Q}_{\qs}(t)$, this yields
\begin{equation}
b_{\qs}(\omega)
=
\frac{c^{\rm Q}_{\qs}(\omega)}
{2\pi\left[n(\omega,T)+1\right]},
\label{eq:app_final_relation}
\end{equation}
which is Eq.~(2) in the main text.

\section{Acknowledgments}

\begin{acknowledgments}
Z.Z. acknowledges support from a Newton International Fellowship from The Royal Society (No.~NIF/R1/254124) and the Oxford ARC system. 
\end{acknowledgments}

\bibliographystyle{apsrev4-2}
\bibliography{cite}

\end{document}